\documentstyle[12pt]{article}
\newlength{\mathspace}
\def\Dbar {\hbox{$D$\kern-0.52em\raise 0.2ex\hbox{/}\kern +0.1em}}
\def\dbar {\hbox{$\partial$\kern-0.52em\raise 0.2ex\hbox{/}\kern +0.1em}}

\begin{document}
\baselineskip=0.7cm
\setlength{\mathspace}{2.5mm}



\begin{titlepage}

    \begin{normalsize}
                
    \end{normalsize}
    \begin{large}
       \vspace{1cm}
       \begin{center}
         {\bf Integrable Models and the Higher Dimensional Representations of
Graded Lie Algebras} 
       \end{center}
    \end{large}

  \vspace{5mm}

\begin{center}
           
             \vspace{.5cm}

            J. C. Brunelli
           \footnote{E-mail address: brunelli@fsc.ufsc.br}\\
Universidade Federal de Santa Catarina\\
 
Departamento de Fisica -- CFM\\
Campus Universitario -- Trindade\\
C. P. 476, CEP 88040-900\\
Florianopolis, SC -- BRAZIL\\
and \\
Ashok Das\\
Department of Physics and Astronomy\\
University of Rochester\\
Rochester, N.Y. 14627, USA\\

      \vspace{2.5cm}

    \begin{large} ABSTRACT \end{large}
        \par
\end{center}
 \begin{normalsize}
\ \ \ \
We construct a zero curvature formulation, in superspace, for the sTB-B
hierarchy which naturally reduces to the zero curvature condition in terms of
components, thus solving one of the puzzling features of this model. This
analysis, further, suggests a systematic method of constructing higher
dimensional representations for the zero curvature condition starting with the
fundamental representation. We illustrate this with the examples of the sTB
hierarchy and the sKdV hierarchy. This would be particularly useful in
constructing explicit higher dimensional representations of graded Lie
algebras. 
\end{normalsize}

\end{titlepage}
\vfil\eject

\begin{large}
\noindent{\bf 1. Introduction:}
\end{large}

\vspace{.5cm}
The supersymmetric integrable models[1-5] have drawn a lot of attention in recent
years for a variety of reasons. One among them is their connection with the
superstring theories through the matrix models. In this connection, it is a
particular supersymmetrization of the bosonic integrable models that has proved
interesting[6]. In a recent paper[7], we studied, in detail, the properties of such a
theory, namely, the sTB-B hierarchy which reduces to all other familiar ()-B
theories under suitable field redefinitions or reductions. One of the puzzling
features to emerge from this study was the fact that the zero curvature
formulation of this hierarchy, in superspace, can be given in terms of $2\times
2$  matrices belonging to $SL(2)\oplus U(1)$ which is related to the second
Hamiltonian structure of the Two Boson (TB) hierarchy. Unlike other
supersymmetric integrable models, however, this does not reduce to the
component zero curvature formulation. In fact, as we showed in [7], the
component zero
curvature formulation of this hierarchy can be given in terms of $4\times 4$
graded matrices belonging to $OSp(2|2)$. Thus, there are two puzzling features. 
First, it is not clear what is the zero curvature condition in superspace which
will naturally reduce to the zero curvature formulation of the components.
Second, neither $SL(2)\oplus U(1)$ nor $OSp(2|2)$ is related to the second
Hamiltonian structure of the hierarchy which, surprisingly, turns out to be
fermionic and, consequently, the relation between the second
Hamiltonian structure and the zero curvature formulation is all the more
puzzling.

We have nothing to say about this second puzzle at this point. However, in this
note, we wish to point out that there is, in fact, a zero curvature formulation
of this hierarchy in superspace which naturally leads to the zero curvature
condition in terms of the components. This analysis, in fact, also enables us
to study higher dimensional representations of the zero curvature conditions
for integrable models. This, in turn, can be used to study the higher 
dimensional
representations of graded Lie algebras. Of course, the general properties of
the representations of graded Lie algebras are quite well known[8-13]. However, this
provides a method for constructing explicit representations which may be
useful. In section 2, we review briefly the known results on the zero curvature
formulation of the sTB-B hierarchy[7] and, then, derive the zero curvature
condition, in superspace, in terms of $4\times 4$ graded matrices belonging to
$OSp(2|2)$ which naturally reduce to the component zero curvature conditions.
In section 3, we give the zero curvature condition for the sTB hierarchy[14] in
terms of $3\times 3$ graded matrices belonging to $OSp(2|2)$ and then explain
how one can systematically construct a $4\times 4$ representation or any higher
representation of the zero curvature condition starting from this. In section 4, 
we describe the construction
of $3\times 3$ as well as $4\times 4$ representations of the zero curvature
condition for the sKdV hierarchy and present a brief conclusion in section 5.
\vspace{.5cm}

\begin{large}
\noindent{\bf 2. Zero Curvature Condition for sTB-B:}
\end{large}

\vspace{.5cm}
The sTB-B hierarchy is described in terms of a Lax operator
\begin{equation}
L = D^{2} - (D\phi_{0}) + D^{-2}(D\phi_{1})\label{A_{1}}
\end{equation}
and a nonstandard Lax equation, in superspace, of the form
\begin{equation}
\frac{\partial L}{\partial t_{n}} = \left[(L^{n})_{\geq 1},
L\right]\label{A_{2}}
\end{equation}
Here $\phi_{0}$ and $\phi_{1}$ are fermionic superfields and
\[
D = \frac{\partial}{\partial\theta} + \theta \frac{\partial}{\partial x}
\]
represents the supercovariant derivative whose square is $\partial_{x}$. (We
refer the readers to [7] for details about the notation and the hierarchy.)

The conventional zero curvature formulation starts from the linear problem
associated with the system given by
\begin{equation}
L \chi  =  (D^{2} - (D\phi_{0}) + D^{-2}(D\phi_{1})) \chi  =  \lambda
\chi\label{A_{3}}
\end{equation}
where $\lambda$ is the spectral parameter. From this, it can be easily shown
that
\begin{eqnarray}
\partial_{x} \left(\begin{array}{c}
\chi\\
(D^{-2}(D\phi_{1})\chi)
\end{array}\right) &  = & {\cal A}_{1} \left(\begin{array}{c}
\chi\\
(D^{-2}(D\phi_{1})\chi)
\end{array}\right)\nonumber\\
 & = &  \left(\begin{array}{cc}
\lambda + (D\phi_{0}) & -1\\
(D\phi_{1}) & 0
\end{array}\right) \left(\begin{array}{c}
\chi\\
(D^{-2}(D\phi_{1})\chi)
\end{array}\right)\label{A_{4}}
\end{eqnarray}
Similarly, we can define another $2\times 2$ matrix
\begin{equation}
{\cal A}_{0} = \left(\begin{array}{cc}
A & B \\
C & E
\end{array}\right)\label{A_{5}}
\end{equation}
associated with the time evolution of the linear wave function, namely,
\[
\partial_{t} \left(\begin{array}{c}
\chi\\
(D^{-2}(D\phi_{1})\chi)
\end{array}\right) = {\cal A}_{0}  \left(\begin{array}{c}
\chi\\
(D^{-2}(D\phi_{1})\chi)
\end{array}\right)
\]
It is now
straightforward to show that the zero curvature condition associated with these
two potentials, namely,
\[
\partial_{t} {\cal A}_{1} - \partial_{x} {\cal A}_{0} - \left[{\cal A}_{0} ,
{\cal A}_{1}\right] = 0
\]
would lead to the sTB-B hierarchy provided
\begin{eqnarray}
A & = & E + B_{x} - (\lambda + (D\phi_{0})) B\nonumber\\
C & = & E_{x} - B (D\phi_{1})\label{A_{6}}
\end{eqnarray}
Furthermore, the dynamical equations describing the hierarchy are given by
\begin{eqnarray}
\frac{\partial (D\phi_{0})}{\partial t} & = & B_{xx} + 2 E_{x} - [B (\lambda +
(D\phi_{0}))]_{x}\nonumber\\
\frac{\partial (D\phi_{1})}{\partial t} & = & E_{xx} - 2 B_{x} (D\phi_{1}) - B
(D^{3}\phi_{1}) + E_{x} (\lambda + (D\phi_{0}))\label{A_{7}}
\end{eqnarray}
The two potentials, ${\cal A}_{0}$ and ${\cal A}_{1}$ clearly belong to
$SL(2)\oplus U(1)$ which is related to the second Hamiltonian structure of the
TB hierarchy. Because of the nature of the ()-B supersymmetrization, it
is clear that a zero curvature formulation would necessarily arise where the
superfield potentials belong to the same symmetry algebra as the bosonic
theory. However, as was already pointed out in [7], these potentials do not lead
to a component zero curvature description in a natural way.

On the other hand, it was already shown in [7] that the component zero curvature
formulation for the hierarchy results from the component potentials of the form
\begin{eqnarray}
{\cal A}_{1} & = & \left(\begin{array}{cccc}
\lambda + J_{0} & -1 & 0 & 0\\
J_{1} & 0 & 0 & 0\\
\psi_{0}' & 0 & \lambda + J_{0} & -1\\
\psi_{1}' & 0 & J_{1} & 0
\end{array}\right)\nonumber\\
{\cal A}_{0} & = & \left(\begin{array}{cccc}
F & G & 0 & 0\\
H & K & 0 & 0\\
P & Q & F & G\\
R & S & H & K
\end{array}\right)\label{A_{8}}
\end{eqnarray}
where the functions satisfy the conditions
\begin{eqnarray}
F & = & K + G_{x} - (\lambda + J_{0}) G\nonumber\\
H & = & K_{x} - G J_{1}\nonumber\\
P & = & S + Q_{x} - (\lambda + J_{0}) Q - \psi_{0}' G\nonumber\\
R & = & S_{x} - J_{1} Q - \psi_{1}' G\label{A_{9}}
\end{eqnarray}
These component potentials, in contrast to the superfield potentials of eqs.
(\ref{A_{4}})-(\ref{A_{5}}), are
graded matrices belonging to the algebra $OSp(2|2)$. The higher dimensional
representation of $OSp(2|2)$ of the form
\[
\left(\begin{array}{cc}
A & 0\\
B & C
\end{array}\right)
\]
where each element is a $2\times 2$ block is quite interesting and has been
described in detail in [12]. Thus, there appears to be
no relation between the zero curvature formulation in superspace and the
component formulation which is one of the puzzles mentioned in the introduction.

We now derive a zero curvature condition associated with sTB-B hierarchy in
superspace which naturally reduces to the component potentials of eqs.
(\ref{A_{7}}) and (\ref{A_{8}}). Let us start with the linear eqution in
(\ref{A_{3}}) and apply $D^{2}$ to it. Then, we will obtain
\begin{equation}
(D^{4} - (\lambda + (D\phi_{0})) D^{2} + (D\phi_{1}) - (D^{3}\phi_{0})) \chi 
= 0\label{A_{10}}
\end{equation}
If we now redefine the superfield wavefunction, $\chi$, as
\begin{equation}
\chi \rightarrow e^{(D^{-2} (\lambda + (D\phi_{0})))} \chi\label{A_{11}}
\end{equation}
then, eq. (\ref{A_{10}}) would become
\begin{equation}
(D^{4} + (\lambda + (D\phi_{0})) D^{2} + (D\phi_{1})) \chi = 0\label{A_{12}}
\end{equation}

>From eq. (\ref{A_{12}}), it is easy to show that
\begin{eqnarray}
\partial_{x} \left(\begin{array}{c}
(D^{3}\chi)\\
(D\chi)\\
(D^{2}\chi)\\
\chi
\end{array}\right) & = & {\cal A}_{1}\left(\begin{array}{c}
(D^{3}\chi)\\
(D\chi)\\
(D^{2}\chi)\\
\chi
\end{array}\right)\nonumber\\
 & = & -\left(\begin{array}{cccc}
\lambda + (D\phi_{0}) & (D\phi_{1}) & (D^{2}\phi_{0}) & (D^{2}\phi_{1})\\
-1 & 0 & 0 & 0\\
0 & 0 & \lambda + (D\phi_{0}) & (D\phi_{1})\\
0 & 0 & -1 & 0
\end{array}\right)\left(\begin{array}{c}
(D^{3}\chi)\\
(D\chi)\\
(D^{2}\chi)\\
\chi
\end{array}\right)\label{A_{13}}
\end{eqnarray}
It is, then, straightforward to show that the graded matrix (associated with
the time evolution of the wave function)
\begin{equation}
{\cal A}_{0} = - \left(\begin{array}{cccc}
A_{1} & A_{2} & (DA_{1}) & (DA_{2})\\
B & E & (DB) & (DE)\\
0 & 0 & A_{1} & A_{2}\\
0 & 0 & B & E
\end{array}\right)\label{A_{14}}
\end{equation}
together with ${\cal A}_{1}$ from eq. (\ref{A_{13}}) would yield the sTB-B
hierarchy of equations in eq. (\ref{A_{7}}), from the zero curvature condition, provided
\begin{eqnarray}
A_{1} & = & E + B_{x} - E (\lambda + (D\phi_{0}))\nonumber\\
A_{2} & = & E_{x} - B (D\phi_{1})\label{A_{15}}
\end{eqnarray}

There are several things to note here. First, the potentials ${\cal A}_{0}$ and
${\cal A}_{1}$ in eqs. (\ref{A_{14}}) and (\ref{A_{13}}) respectively are
graded matrices belonging to $OSp(2|2)$ exactly of the same form as in eq.
(\ref{A_{8}}).
Second, if we expand these as
\begin{equation}
{\cal A}_{0,1} = a_{0,1} + \theta \tilde{a}_{0,1}\label{A_{16}}
\end{equation}
then, it is obvious that $-a_{0,1}^{T}$ corresponds to the component potentials
of the zero curvature condition in eqs. (\ref{A_{8}}). (Here $T$ stands for    
the transpose
of the matrix.) This, therefore, completes the derivation of a zero curvature
formulation, in superspace, of the sTB-B hierarchy which naturally reduces to
the component zero curvature condition. However, the other issue of how the
fermionic Hamiltonian structures of this hierarchy are related to the symmetry
algebra $OSp(2|2)$ remains unclear at this point.  
\vspace{.5cm}

\begin{large}
\noindent{\bf 3. Higher Dimensional Representation for sTB:}
\end{large}

\vspace{.5cm}
Normally, all the zero curvature formulations of
integrable models that we know of are in the fundamental representation of the
symmetry algebra. However, the analysis of the previous section is very
interesting from a completely different point of view, namely, it shows how one
can construct, systematically, higher order representations of the zero
curvature condition starting from that in the fundamental  representation.
This, in turn, can be quite useful, particularly in the case of supersymmetric
systems, since it allows us to construct explicit higher dimensional
representations of graded Lie algebras. We will discuss this with some
examples.

First, let us consider the sTB hierarchy[4] described by the Lax operator
\begin{equation}
L = D^{2} - (D\phi_{0}) + D^{-1} \phi_{1}\label{B_{1}}
\end{equation}
and the nonstandard Lax equation
\begin{equation}
\frac{\partial L}{\partial t_{n}} = \left[ (L^{n})_{\geq 1} ,
L\right]\label{B_{2}}
\end{equation}
The linear problem, associated with this system, is given by
\begin{equation}
L \chi = (D^{2} - (D\phi_{0}) + D^{-1}\phi_{1}) \chi = \lambda
\chi\label{B_{3}}
\end{equation}
Operating with $D$ on eq. (\ref{B_{3}}), we obtain
\begin{equation}
(D^{3} - (\lambda + (D\phi_{0})) D + \phi_{1} - (D^{2}\phi_{0})) \chi =
0\label{B_{4}}
\end{equation}

Let us next define the variables
\begin{eqnarray}
\chi_{1} & = & \chi\nonumber\\
\chi_{2} & = & (D\chi_{1}) = (D\chi)\nonumber\\
\chi_{3} & = & (D\chi_{2}) = (D^{2}\chi)\label{B_{5}}
\end{eqnarray}
In terms of these variables, the linear equation in (\ref{B_{4}}) becomes
\begin{equation}
(D\chi_{3}) - (\lambda + (D\phi_{0})) \chi_{2} + (\phi_{1} - (D^{2}\phi_{0}))
\chi_{1} = 0\label{B_{6}}
\end{equation}
Furthermore, it is easy to obtain from eqs. (\ref{B_{5}}) and (\ref{B_{6}})
that
\begin{eqnarray}
\partial_{x} \left(\begin{array}{c}
\chi_{3}\\
\chi_{1}\\
\chi_{2}
\end{array}\right) & =  & {\cal A}_{1} \left(\begin{array}{c}
\chi_{3}\\
\chi_{1}\\
\chi_{2}
\end{array}\right)\nonumber\\
 & = &  \left(\begin{array}{ccc}
\lambda + (D\phi_{0}) & -(D\phi_{1}) + (D^{3}\phi_{0}) & \phi_{1}\\
1 & 0 & 0\\
0 & -\phi_{1} + (D^{2}\phi_{0}) & \lambda + (D\phi_{0})
\end{array}\right) \left(\begin{array}{c}
\chi_{3}\\
\chi_{1}\\
\chi_{2}
\end{array}\right)\label{B_{7}}
\end{eqnarray}
If we now define another graded matrix of the form (associated with the time
evolution of the wave function)
\begin{equation}
{\cal A}_{0} = \left(\begin{array}{ccc}
A & B & C\\
E & F & G\\
H & J & (A+F)
\end{array}\right)\label{B_{8}}
\end{equation}
then, it is straightforward to check that the zero curvature condition
\[
\partial_{t} {\cal A}_{1} - \partial_{x} {\cal A}_{0} - \left[{\cal A}_{0} ,
{\cal A}_{1}\right] = 0
\]
leads to the sTB hierarchy of equations provided
\begin{eqnarray}
F & = & (DG) - E_{x}\nonumber\\
H & = & - G + (DE)\nonumber\\
A & = &  (DG)  + E (\lambda + (D\phi_{0}))\nonumber\\
B & = & (DG_{x}) - E_{xx} + E (-(D\phi_{1}) + (D^{3}\phi_{0})) + G (- \phi_{1}
+ (D^{2}\phi_{0}))\nonumber\\
J & = & G_{x} - (DE_{x}) + E (- \phi_{1} + (D^{2}\phi_{0}))\nonumber\\
C & = & G_{x} + E \phi_{1} + G (\lambda + (D\phi_{0}))\label{B_{9}}
\end{eqnarray} 
The dynamical equations of the hierarchy are given by
\begin{eqnarray}
\frac{\partial (D\phi_{0})}{\partial t} & = & \left[2(DG) - E_{x} + E (\lambda
+ (D\phi_{0}))\right]_{x} + \phi_{1} (2G - (DE)) - \phi_{0,x} G\nonumber\\
\frac{\partial \phi_{1}}{\partial t} & = & G_{xx} + (E \phi_{1})_{x} + G_{x}
(\lambda + (D\phi_{0})) + (D \phi_{1}G) + E_{x} \phi_{1}\label{B_{10}}
\end{eqnarray}

This provides a zero curvature formulation of the sTB hierarchy in terms of
potentials belonging to the fundamental representation of $OSp(2|2)$. We note
here that the zero curvature formulation of sTB is already known[14]. However, we
have derived it differently here which readily lends itself to a higher
dimensional generalization. We also note here that the sTB hierarchy is known
to reduce to the sKdV hierarchy when $\lambda = 0$ and $\phi_{0}=0$. Imposing 
this consistently,
we note from eq. (\ref{B_{10}}) that we must have, in this case,
\begin{equation}
G = \frac{1}{2} (DE)\label{B_{11}}
\end{equation}
in addition to the other conditions in eq. (\ref{B_{9}}).

In trying to generalize the zero curvature to the next higher dimensional
representation of $OSp(2|2)$, namely, the $4\times 4$ representation, we
define, in addition to the relations in eq. (\ref{B_{5}}),
\begin{equation}
\chi_{4} = (D\chi_{3}) = (D^{3}\chi)\label{B_{12}}
\end{equation}
In this case, the linear equation, (\ref{B_{6}}), becomes
\begin{equation}
\chi_{4} = (\lambda + (D\phi_{0})) \chi_{2} + (-\phi + (D^{2} \phi_{0}))
\chi_{1}\label{B_{13}}
\end{equation}
>From eqs. (\ref{B_{5}}), (\ref{B_{12}}) and (\ref{B_{13}}), it is easy to
obtain that
\[
\partial_{x} \left(\begin{array}{c}
\chi_{3}\\
\chi_{1}\\
\chi_{2}\\
\chi_{4}
\end{array}\right) =  \tilde{\cal A}_{1} \left(\begin{array}{c}
\chi_{3}\\
\chi_{1}\\
\chi_{2}\\
\chi_{4}
\end{array}\right)
\]
where
\begin{equation}
\tilde{\cal A}_{1} =  \left(\begin{array}{cccc}
\lambda + (D\phi_{0}) & -(D\phi_{1}) + (D^{3}\phi_{0}) & \phi_{1} & 0\\
1 & 0 & 0 & 0\\
0 & -\phi_{1} + (D^{2}\phi_{0}) & \lambda + (D\phi_{0}) & 0\\
\begin{array}{c}
-\phi_{1} + (D^{2}\phi_{0})\\
  
\end{array} & \begin{array}{c}
                              (\lambda + (D\phi_{0}))(-\phi_{1} +
                                  (D^{2}\phi_{0}))\\
                               + (-(D^{2}\phi_{1}) + (D^{4}\phi_{0}))
                              \end{array} & \begin{array}{c}
                                             (\lambda + (D\phi_{0}))^{2}\\
                                                + (D^{3}\phi_{0})
                                             \end{array} & 
\begin{array}{c}
0\\
  
\end{array}
\end{array}\right)\label{B_{14}}
\end{equation}

>From eq. (\ref{B_{14}}), we note that the first $3\times 3$ block of this
matrix is the same as ${\cal A}_{1}$ of eq. (\ref{B_{7}}). Let us next define
\begin{equation}
\tilde{\cal A}_{0} = \left(\begin{array}{cccc}
A & B & C & 0\\
E & F & G & 0\\
H & J & (A+F) & 0\\
M & N & P & 0
\end{array}\right)\label{B_{15}}
\end{equation}
where, once again, the first $3\times 3$ block is the same as ${\cal A}_{0}$ in
eq. (\ref{B_{8}}). The structures of the matrices in eqs.
(\ref{B_{14}}) and (\ref{B_{15}}) guarantee that the dynamical equations of 
(\ref{B_{10}}) would automatically follow with the same conditions as in eq.
(\ref{B_{9}}) from the first $3\times 3$ block of these matrices. The          
additional conditions coming from the time evolution of the
elements in the fourth row only provide  consistency conditions and determine
\begin{eqnarray}
M & = & (\lambda + (D\phi_{0})) H + E (-\phi_{1} + (D^{2}\phi_{0}))\nonumber\\
N & = & J_{x} + (A+F) (-\phi_{1} + (D^{2}\phi_{0})) + H (-(D\phi_{1}) +
(D^{3}\phi_{0}))\nonumber\\
P & = & A_{x} + B - E (-(D\phi_{1}) + (D^{3}\phi_{0})) - \phi_{1} H + (A+F)
(\lambda + (D\phi_{0}))\nonumber\\
 &  &  - G (-\phi_{1} + (D^{2}\phi_{0}))\label{B_{16}}
\end{eqnarray}
These consistency conditions can also be derived by analyzing the time
evolution equation for the wave function of the linear equation. This,
therefore, gives a zero curvature formulation of the sTB hierarchy for
potentials belonging to the $4$-dimensional representation of $OSp(2|2)$ and it
is clear that this method can be generalized to obtain the zero curvature
formulation for any higher dimensional representation.
\vspace{.5cm}

\begin{large}
\noindent{\bf 4. Higher Dimensional Representation for sKdV:}
\end{large}

\vspace{.5cm}
The second Hamiltonian structure of the sKdV hierarchy is known[3] to be the
superconformal algebra. Correspondingly, the zero curvature formulation of the
sKdV hierarchy is known in terms of potentials belonging to $OSp(2|1)$[14-15]. In
this section, we will describe how the zero curvature formulation can be
extended to any higher dimensional representation of $OSp(2|1)$. First, let us
review briefly the standard zero curvature formulation in a way that will
generalize to higher dimensional representation easily.

The sKdV hierarchy is described in terms of the Lax operator
\begin{equation}
L = D^{4} + D \phi\label{C_{1}}
\end{equation}
and the standard Lax equation
\begin{equation}
\frac{\partial L}{\partial t_{n}} = \left[(L^{\frac{2n+1}{2}})_{+} ,
L\right]\label{C_{2}}
\end{equation}
The linear equation associated with this hierarchy is given by
\begin{equation}
L \chi = (D^{4} + D \phi) \chi = 0\label{C_{3}}
\end{equation}
This can also be equivalently written as
\begin{equation}
(D^{3} + \phi) \chi = 0\label{C_{4}}
\end{equation}

Let us now define, as in the previous section,
\begin{eqnarray}
\chi_{1} & = & \chi\nonumber\\
\chi_{2} & = & (D\chi_{1}) = (D\chi)\nonumber\\
\chi_{3} & = & (D\chi_{2}) = (D^{2}\chi)\label{C_{5}}
\end{eqnarray}
Then, the linear equation of eq. (\ref{C_{4}}) can be written as
\begin{equation}
(D\chi_{3}) + \phi \chi_{1} = 0\label{C_{6}}
\end{equation}
>From this, we can easily show that
\begin{equation}
\partial_{x} \left(\begin{array}{c}
\chi_{3}\\
\chi_{1}\\
\chi_{2}
\end{array}\right) = {\cal A}_{1} \left(\begin{array}{c}
\chi_{3}\\
\chi_{1}\\
\chi_{2}
\end{array}\right) = \left(\begin{array}{ccc}
0 & - (D\phi) & \phi\\
1 & 0 & 0\\
0 & - \phi & 0
\end{array}\right) \left(\begin{array}{c}
\chi_{3}\\
\chi_{1}\\
\chi_{2}
\end{array}\right)\label{C_{7}}
\end{equation}
As before, let us define the matrix ${\cal A}_{0}$ associated with the time
evolution of the wave function of the linear equation as
\begin{equation}
{\cal A}_{0} = \left(\begin{array}{ccc}
A & B & C\\
E & F & G\\
H & J & (A+F)
\end{array}\right)\label{C_{8}}
\end{equation}
Then, it is straightforward to show that the zero curvature condition
\[
\partial_{t} {\cal A}_{1} - \partial_{x} {\cal A}_{0} - \left[{\cal A}_{0} ,
{\cal A}_{1}\right] = 0
\]
leads to the sKdV hierarchy of equations provided
\begin{eqnarray}
G & = & H = \frac{1}{2} (DE)\nonumber\\
C & = & - J = \frac{1}{2} (DE_{x}) + E \phi\nonumber\\
A & = & - F = \frac{1}{2} E_{x}\nonumber\\
B & = & - \frac{1}{2} E_{xx} - E (D\phi) - \frac{1}{2} (DE) \phi\label{C_{9}}
\end{eqnarray}
Furthermore, the dynamical equation of the hierarchy is given by
\begin{equation}
\frac{\partial \phi}{\partial t} = \frac{1}{2} (DE_{xx}) + (E \phi)_{x} +
\frac{1}{2} (DE)(D\phi) + \frac{1}{2} E_{x} \phi\label{C_{10}}
\end{equation}

This gives the zero curvature formulation of the sKdV hierarchy in the
fundamental representation of $OSp(2|1)$ and it can be compared with eq.
(\ref{B_{10}})
with proper identifications. To go beyond the fundamental representation, we
again define, as before,
\begin{equation}
\chi_{4} = (D\chi_{3}) = (D^{3}\chi)\label{C_{11}}
\end{equation}
so that the linear equation, (\ref{C_{6}}), becomes
\begin{equation}
\chi_{4} = - \phi \chi_{1}\label{C_{12}}
\end{equation}
It now follows from eqs. (\ref{C_{5}}), (\ref{C_{11}}) and (\ref{C_{12}}) that
\begin{equation}
\partial_{x} \left(\begin{array}{c}
\chi_{3}\\
\chi_{1}\\
\chi_{2}\\
\chi_{4}
\end{array}\right) = \tilde{\cal A}_{1} \left(\begin{array}{c}
\chi_{3}\\
\chi_{1}\\
\chi_{2}\\
\chi_{4}
\end{array}\right) = \left(\begin{array}{cccc}
0 & - (D\phi) & \phi & 0\\
1 & 0 & 0 & 0\\
0 & -\phi & 0 & 0\\
-\phi & - (D^{2}\phi) & 0 & 0
\end{array}\right)  \left(\begin{array}{c}
\chi_{3}\\
\chi_{1}\\
\chi_{2}\\
\chi_{4}
\end{array}\right)\label{C_{13}}
\end{equation}
Once again, we note that the first $3\times 3$ block is nothing other than
${\cal A}_{1}$. We can again define
\begin{equation}
\tilde{\cal A}_{0} = \left(\begin{array}{cccc}
A & B & C & 0\\
E & F & G & 0\\
H & J & (A+F) & 0\\
M & N & P & 0
\end{array}\right)\label{C_{14}}
\end{equation}
where the first $3\times 3$ block is ${\cal A}_{0}$. From the structure of the
matrices in eqs. (\ref{C_{13}}) and (\ref{C_{14}}), we note that the dynamical
equation (\ref{C_{10}}) is guaranteed with the identifications in eq.
(\ref{C_{9}}) from the first $3\times 3$ block. The additional equations       
arising from the elements in the fourth
row only provide consistency conditions and determine
\begin{eqnarray}
M & = & - E \phi\nonumber\\
N & = & - C_{x} - G (D\phi)\nonumber\\
P & = & G \phi\label{C_{15}}
\end{eqnarray}
which can also be determined from an analysis of the time evolution of the wave
function associated with the linear equation.

This provides a generalization of the zero curvature formulation for the sKdV
hierarchy to $4$-dimensions. This can also be compared with the reduction
resulting
from the four dimensional representation of the sTB hierarchy of
the previous section. We note here that $OSp(2|1)$ does not have an irreducible
representation of dimension $4$. This, therefore, appears to correspond to a
reducible representation of $OSp(2|1)$ for $j = 0 + \frac{1}{2}$. However, the
five dimensional, adjoint representation is not hard to construct as the
discussion shows. 
\vspace{.5cm}

\begin{large}
\noindent{\bf 6. Conclusion:}
\end{large}

\vspace{.5cm}
We have obtained  a zero curvature formulation of the sTB-B hierarchy, in
superspace, which naturally reduces to the component zero curvature
formulation. This solves one of the puzzling features found in connection with
this hierarchy. Our analysis further suggests a systematic way of generating
the zero curvature formulation to any higher dimensional representation of the
symmetry algebra for any integrable hierarchy. This, in turn, may be quite
useful in constructing explicit higher dimensional representations of graded
Lie algebras. We have shown how the method works for the sTB and the sKdV
hierarchies.

One of us (AD) would like to thank Profs. H. Aratyn and A. B. Balantekin for
useful discussions on the representations of the graded Lie algebras.
This work was supported in part by US Department of Energy Grant 
No. DE-FG02-91ER40685 and NSF-INT-9602559. JCB is supported by CNPq, Brazil. 
\vspace{.5cm}

\begin{large}
\noindent{\bf References:}
\end{large}

\vspace{.5cm}
\begin{enumerate}
\item{ }B. A. Kupershmidt, {\it Elements of Super Integrable Systems: basic
techniques and results}, Kluwer Acad. Publ. (1987).
\item{ }Y. Manin and A. O. Radul, Comm. Math. Phys. {\bf 98}, 65 (1985).
\item{ }P. Mathieu, J. Math. Phys. {\bf 29}, 2499 (1988).
\item{ }J. C. Brunelli and A. Das, Phys. Lett. {\bf 337B}, 303 (1994); Int. J.
Mod. Phys. {\bf A10}, 4563 (1995).
\item{ }H. Aratyn and A. Das, \lq\lq The sAKNS Hierarchy", submitted to Phys.
Lett. A.
\item{ }K. Becker and M. Becker, Mod. Phys. Lett. {\bf A8}, 1205 (1993); J. M.
Figueroa-O'Farril and S. Stanciu, Phys. Lett. {\bf 316B}, 282 (1993).
\item{ }J. C. Brunelli and A. Das, Phys. Lett. {\bf 409B}, 229 (1997).
\item{ }A. Pais and V. Rittenberg, J. Math. Phys. {\bf 16}, 2062 (1975).
\item{ }M. Scheunert, W. Nahm and V. Rittenberg, J. Math. Phys. {\bf 18}, 146
(1977); {\it ibid} {\bf 18}, 155 (1977).
\item{ }J. W. B. Hughes, J. Math. Phys. {\bf 22}, 245 (1981).
\item{ }V. G. Kac in, {\it Differential Geometric Methods in Mathematical
Physics}, ed. K. Bleuler, H. R. Petry and A. Reetz (Springer, Berlin 1987).
\item{ }A. B. Balantekin and I. Bars, J. Math. Phys. {\bf 22}, 1149 (1981); {\it
ibid} {\bf 22}, 1810 (1981).
\item{ }L. Frappat, P. Sorba and A. Sciarrino, {\it Dictionary on Lie
Superalgebras}, hep-th/9607161.
\item{ }H. Aratyn, A. Das, C. Rasinariu and A. H. Zimerman, to be published in
Lecture Notes in Physics.
\item{ }A. Das and S. Roy, J. Math. Phys. {\bf 31}, 2145 (1990).
\end{enumerate}

\vfil
\eject 

\end{document}